\documentstyle[aps,epsf,floats]{revtex}

\tighten
\newcommand{\ph}{{\rule{0mm}{3mm}}}
\newcommand{\slsh}[1]{\mbox{$\not\! #1$}}
\newcommand{\bm}[1]{\bbox{#1}}
\newcommand{\simorder}{\raisebox{-4pt}{$\, \stackrel{\textstyle >}{\sim} \,$}}

\begin{document}
 
\draft
\title{
\begin{flushright}
\begin{minipage}{3 cm}
\small
hep-ph/9711488\\
NIKHEF 97-046\\
VUTH 97-19\\
FNT/T-97/14
\end{minipage}
\end{flushright}
Leading asymmetries in two-hadron production\\
in $e^+e^-$ annihilation at the $Z$ pole}

\author{D. Boer$^1$, R. Jakob$^2$ and P.J. Mulders$^{1,3}$}
\address{\mbox{}\\
$^1$National Institute for Nuclear Physics and High--Energy
Physics (NIKHEF)\\	
P.O. Box 41882, NL-1009 DB Amsterdam, the Netherlands\\
\mbox{}\\
$^2$Universit\`{a} di Pavia and INFN, Sezione di Pavia\\ 
Via Bassi 6, 27100 Pavia, Italy\\
\mbox{}\\
$^3$Department of Physics and Astronomy, Free University \\
De Boelelaan 1081, NL-1081 HV Amsterdam, the Netherlands
}

\maketitle
\begin{center}November, 1997\end{center}

\begin{abstract}
We present the leading unpolarized and single spin asymmetries in inclusive
two-hadron production in electron-positron annihilation at the  
$Z$ pole.  
The azimuthal dependence in the unpolarized 
differential cross section of almost
back-to-back hadrons is a leading $\cos(2\phi)$ asymmetry, which arises 
solely due to
the intrinsic transverse momenta of the quarks. 
An extensive discussion on how to measure 
this asymmetry and the accompanying time-reversal odd fragmentation 
functions is given. A simple estimate 
indicates that the asymmetry could be of the order of a percent. 
\end{abstract}


Recently, we have presented the results of the complete tree-level 
calculation of inclusive two-hadron production in electron-positron 
annihilation via one photon up to subleading order in $1/Q$~\cite{Boer},
where the scale $Q$ is defined by the (timelike) photon momentum $q$ (with
$Q^2 \equiv q^2$) and given by $Q$ = $\sqrt{s}$. 
The quantity $Q$ had to be much larger than characteristic
hadronic scales, but -- being interested in effects at subleading order -- we
considered energies only well below the threshold for the production of $Z$
bosons.  

In this article we extend those results to electron-positron 
annihilation into a $Z$ boson, such that the results can be used to analyze
LEP-I data. We will neglect contributions from photon exchange and
$\gamma$-$Z$ interference terms, which are known to be numerically irrelevant 
on 
the $Z$ pole. Only leading order $(1/Q)^0$ effects are discussed, since for
$Q \simorder M_Z$ the power corrections of order $1/Q$ are expected to be 
completely negligible. Furthermore, we will focus on tree 
level, i.e., order $(\alpha_s)^0$. A rich structure nevertheless arises when
taking into account the intrinsic transverse momentum of the quarks and,
possibly, polarization of the detected hadrons in the final
state. By accounting for intrinsic transverse momentum effects we extend the
results in the analysis of Chen {\em et al.\/}~\cite{Chen-et-al-95}, where no
azimuthal asymmetries arising from transverse momenta have been considered.

For details of the calculation and the formalism we refer to \cite{Boer}. We 
shortly repeat the essentials. We consider the process 
$e^-+e^+\rightarrow {\rm hadrons}$, where the two leptons
(with momentum $l$ for the $e^-$ and $l^\prime$ for the $e^+$, 
respectively)  annihilate into a $Z$ boson with 
momentum $q = l + l^\prime$, which is timelike with 
$q^2 \equiv Q^2$. Denoting the momentum of the two outgoing hadrons by $P_h$
($h$ = 1, 2) we use invariants $z_h$ = $2P_h\cdot q/Q^2$. 
We will consider the case where the two hadronic momenta $P_1$ and $P_2$ 
do not belong to the same jet (i.e., $P_1\cdot P_2$ is of order $Q^2$). 
In principle, the momenta can also be considered as the jet momenta
themselves, but then effects due to intrinsic transverse momentum 
will be absent. 
We will treat the production of hadrons
of which the spin states are  
characterized by a spin vector $S_h$ ($h$ = 1, 2), satisfying 
$P_h\cdot S_h = 0$ and $-1\leq S_h^2 \leq 0$. 
In this way we can treat the case of 
unpolarized final states or final state hadrons with spin-0 and spin-1/2.
In the present article we will 
disregard the polarization of hadron two (summation over spins).
The final states which have to be identified and analyzed for the effects we
discuss are simpler than the ones investigated by Artru and 
Collins \cite{Artru-Collins-96}, who proposed to measure azimuthal 
correlations in four-hadron production.

The cross section (including a factor 1/2 from averaging over incoming
polarizations) for two-particle inclusive $e^+e^-$ 
annihilation is given by 
\begin{equation}
\frac{P_1^0 \,P_2^0\,\,d\sigma^{(e^+e^-)}}{d^3P_1\,d^3P_2}
=\frac{\alpha_w^2}{4\,\left((Q^2 - M_Z^2)^2+\Gamma_Z^2M_Z^2\right)\, Q^2} 
L_{\mu\nu}{\cal W}^{\mu\nu},
\label{cross2}
\end{equation}
with $\alpha_w = e^2/ (16 \pi \sin^2 \theta_W \cos^2 \theta_W)$ and the 
helicity-conserving lepton tensor (neglecting the lepton masses and
polarization) is given by 
\begin{equation} \label{leptten2}
L_{\mu\nu} (l, l^\prime)
=  (g_V^l{}^2 + g_A^l{}^2) \left[ 2 l_\mu l^\prime_\nu
+ 2 l_\nu l^\prime_\mu - Q^2 g_{\mu\nu} \right]
-(2 g_V^l g_A^l) 2i 
\,\epsilon_{\mu\nu\rho\sigma} l^\rho l^{\prime \sigma},
\end{equation}
where $g_V^l$, $g_A^l$ denote the vector and axial-vector
couplings of the $Z$ boson to the leptons, respectively, 
and the hadron tensor is given by
\begin{equation}
{\cal W}_{\mu\nu}( q;  P_1 S_1; P_2 S_2 ) =
\int \frac{d^3 P_X}{(2\pi)^3 2P_X^0}
\delta^4 (q-P_X - P_1 - P_2)
H_{\mu\nu}(P_X;  P_1 S_1; P_2 S_2),
\label{hadrten2}
\end{equation}
with 
\begin{equation}
H_{\mu\nu}(P_X;  P_1 S_1; P_2 S_2)
= \langle 0 |J_\mu (0)|P_X; P_1 S_1; P_2 S_2 \rangle
\langle P_X; P_1 S_1; P_2 S_2 |J_\nu (0)| 0 \rangle ,
\end{equation}
where a summation over spins of the unobserved {\em out}-state $X$ 
is understood.

\begin{figure}[htb]
\begin{center}
\leavevmode \epsfxsize=10cm \epsfbox{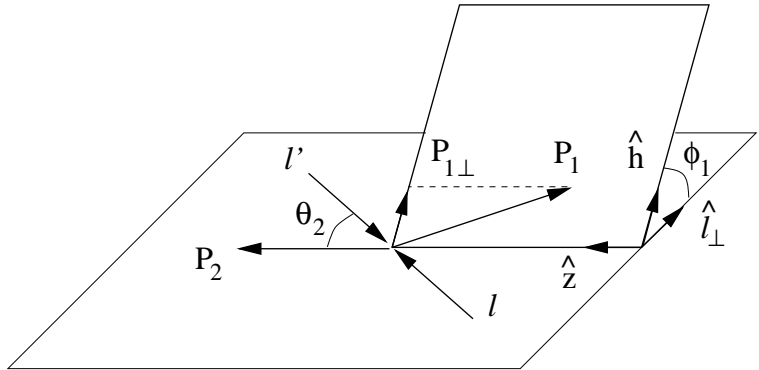}
\vspace{0.2 cm}
\caption{\label{fig:kinann} Kinematics of the annihilation process in
the lepton center of mass frame for a back-to-back jet situation.
$P_1$ ($P_2$) is the momentum of a fast hadron in jet one (two).}
\end{center}
\end{figure}

In order to expand the lepton and hadron tensors in terms of independent
Lorentz structures, it is convenient to
work with vectors orthogonal to $q$. 
A normalized timelike vector $\hat t$ is defined 
by the boson momentum $q$ and
a normalized spacelike vector $\hat z$ is defined by $\tilde P^\mu$ = 
$P^\mu - (P\cdot q/q^2)\,q^\mu$ for one of the outgoing momenta, say $P_2$,
\begin{eqnarray}
\hat t^\mu& \equiv & \frac{q^\mu}{Q},\\
\hat z^\mu & \equiv &
\frac{Q}{P_2\cdot q}\,\tilde P^\mu_2
\ =\   2\,\frac{P_2^\mu}{z_2 Q} - \frac{q^\mu}{Q}. 
\end{eqnarray}
Vectors orthogonal to $\hat t$ and $\hat z$ are obtained with help of 
the tensors
\begin{eqnarray}
&& g_{\perp}^{\mu\nu}
\equiv  g^{\mu\nu} -\hat t^\mu \hat t^\nu +
\hat z^\mu \hat z^\nu, \\
&& \epsilon_\perp^{\mu \nu} \equiv
-\epsilon^{\mu \nu \rho \sigma} \hat t_\rho \hat z_\sigma
\ =\ \frac{1}{(P_2\cdot q)}\,\epsilon^{\mu \nu \rho
\sigma} P_{2\,\rho}q_\sigma.
\end{eqnarray}
Since we have chosen hadron two 
to define the longitudinal direction, the momentum
of hadron one can be used to express the directions orthogonal to 
$\hat t$ and $\hat z$. We define 
the normalized vector $\hat h^\mu$ = 
$P_{1\perp}^\mu/\vert \bm P_{1\perp} \vert$, with 
$P_{1\perp}^\mu$ = $g_\perp^{\mu\nu}\,P_{1\nu}$, and the second
orthogonal direction is given by  
$\epsilon_\perp^{\mu \nu} \hat h_\nu$ (see Fig.\ \ref{fig:kinann}). 
We use boldface vectors to denote the two-dimensional Euclidean part
of a four-vector, such that 
$P_{1\perp}\cdot P_{1\perp}=-\bm P_{1\perp}\cdot \bm P_{1\perp}$.

In the calculation of the hadron tensor it will be convenient to define 
{\em lightlike} directions using the
hadronic (or jet) momenta. 
The momenta can then be parametrized (remember that $P_1\cdot P_2$ is of
order $Q^2$) using dimensionless 
lightlike vectors
$n_+$ and $n_-$ satisfying $n_+^2 = n_-^2 = 0$ and $n_+\cdot n_-$ = 1,
\begin{eqnarray}
&& P_1^\mu \equiv \frac{z_1 Q}{\sqrt{2}}\,n_-^\mu,\\
&& P_2^\mu \equiv \frac{z_2 Q}{\sqrt{2}}\,n_+^\mu,\\
&&q^\mu \equiv \frac{ Q}{\sqrt{2}}\,n_-^\mu
+ \frac{ Q}{\sqrt{2}}\,n_+^\mu + q_T^\mu,
\end{eqnarray}
with $q_T^2 \equiv -Q_T^2$. We have neglected hadron mass terms and
considering the case
of two back-to-back jets we have $Q_T^2 \ll Q^2$. 
We will use the 
notation $p^\pm = p \cdot n_\mp$ for a generic momentum 
$p$. As 
momentum $P_2$ defines the vector $\hat z^\mu$, 
\begin{equation}
P_{1\perp}^\mu = - z_1\,q_T^\mu = - z_1\, Q_T\, \hat h^\mu.
\end{equation}
Vectors transverse to $n_+$ and $n_-$ one obtains using the tensors
\begin{eqnarray}
&&g^{\mu\nu}_T \ \equiv \ g^{\mu\nu}
- n_+^{\,\{\mu} n_-^{\nu\}}, \\
&&\epsilon^{\mu\nu}_T \ \equiv
\ \epsilon^{\mu\nu\rho\sigma} n_{+\rho}n_{-\sigma},
\end{eqnarray}
where the brackets around the indices indicate symmetrization. 
The lightlike directions can easily be expressed in $\hat t$,
$\hat z$ and a perpendicular vector,
\begin{eqnarray}
n_+^\mu & = & \frac{1}{\sqrt{2}} \left[ \hat t^\mu + \hat z^\mu \right],
\label{transverse1} \\
n_-^\mu & = & 
\frac{1}{\sqrt{2}} \left[ \hat t^\mu - \hat z^\mu
+ 2\,\frac{Q_T^{}}{Q}\,\hat h^\mu \right], 
\label{transverse2}
\end{eqnarray}
showing that the differences between $g^{\mu\nu}_\perp$ and $g^{\mu\nu}_T$ 
are of order $1/Q$. We will see however that 
taking transverse momentum into account does not automatically lead to 
suppression. 

To leading order 
the expression for the hadron tensor, including quarks and antiquarks, is
\begin{eqnarray}
{\cal W}^{\mu\nu}&=&3 \int dp^- dk^+ d^2\bm{p}_T^{} d^2 
\bm{k}_T^{}\, \delta^2(\bm{p}_T^{}+
\bm{k}_T^{}-\bm{q}_T^{})\, 
\left. \text{Tr}\left( \overline \Delta (p) 
V^\mu \Delta 
(k) V^\nu \right)\right|_{p^+\,k^-} + \left(\begin{array}{c} 
q\leftrightarrow -q \\ \mu \leftrightarrow \nu
\end{array} \right),
\label{Wmunustart}
\end{eqnarray}
where $V^\mu = g_V \gamma^\mu + g_A \gamma_5 \gamma^\mu$ is the $Z$
boson-quark vertex. We have omitted flavor indices and summation. The 
correlation functions $\Delta$ and $\overline \Delta$ are given by
\cite{Collins-Soper-82}:
\begin{eqnarray}
\Delta_{ij}(k) & = & \sum_X \frac{1}{(2\pi)^4}\int d^4x\ e^{ik\cdot 
x}
\,
\langle 0 \vert \psi_i(x) \vert P_1,S_1; X \rangle
\langle P_1, S_1; X \vert \overline \psi_j(0) \vert 0 \rangle,\\
\overline \Delta_{ij}(p) & = & \sum_X \frac{1}{(2\pi)^4}
\int d^4x\ e^{-ip\cdot x}\,
\langle 0 \vert \overline \psi_j(0) \vert P_2, S_2; X \rangle
\langle P_2,S_2; X \vert \psi_i(x) \vert 0 \rangle
\end{eqnarray}
and the quark momentum $k$ (and similarly for $p$) and the polarization 
vector $S_1$ (from now on we omit $S_2$) are decomposed as follows:
\begin{eqnarray}
k &\approx & \frac{1}{z} P_1 + k_T^{},\\
S_1 & \approx & \frac{\lambda_1}{M_1} P_1 + S_{1T}^{}.
\end{eqnarray}
To leading order in $1/Q$ one has that $z=z_1$. 
The (partly integrated) correlation function $\Delta$ is parametrized as:
\begin{eqnarray}
&&\left. \frac{1}{z} \int dk^+\ \Delta(P_1,S_1;k)
\right|_{k^- = P_1^-/z,\ \bm{k}_{\scriptscriptstyle T}} =
\frac{M_1}{P_1^-} \Biggl\{
D_1\,\frac{\slsh{\!P_1}}{M_1}
+ D_{1T}^\perp\, \frac{\epsilon_{\mu \nu \rho \sigma}
\gamma^\mu P_1^\nu k_T^\rho S_{1T}^\sigma}{M_1^2}
- G_{1s}\,\frac{\slsh{\!P_1} \gamma_5}{M_1}
\nonumber \\[3mm] & &
\qquad 
- H_{1T}\,\frac{i\sigma_{\mu\nu}\gamma_5\,S_{1T}^\mu P_1^\nu}{M_1}
- H_{1s}^\perp\,
\frac{i\sigma_{\mu\nu}\gamma_5\, k_T^\mu P_1^\nu}{M_1^2}
+ H_{1}^\perp\,\frac{\sigma_{\mu \nu} k_T^\mu P_1^\nu}{M_1^2}
\Biggr\},
\label{Deltaexp}
\end{eqnarray}
where the shorthand notation $G_{1s}$ (and similarly for $H_{1s}^\perp$)
stands for the combination
\begin{equation}
G_{1s}(z,\bm{k}_T^{}) = \lambda_1\,G_{1L}
+ G_{1T}\,\frac{(\bm{k}_T^{}\cdot \bm{S}_{1T}^{})}{M_1}.
\end{equation}
We parametrize the antiquark correlation function $\overline \Delta$ in the
same way, except that the distribution functions are 
overlined and the obvious replacements of momenta are done. 

The functions $D_1, \ldots$ in Eq.\ (\ref{Deltaexp}) and $G_{1L}, G_{1T},
\ldots$ in $G_{1s}, \ldots$ are called 
fragmentation functions. One wants to express
the fragmentation functions in terms of the hadron momentum, hence, the
arguments of
the fragmentation functions are chosen to be the lightcone (momentum)
fraction $z= P_1^-/k^-$ of the produced hadron with respect to the
fragmenting quark and $\bm{k}_T^\prime \equiv -z\bm{k}_T^{}$, which is the 
transverse 
momentum of the hadron in a frame where the quark has no transverse momentum.
In order to switch from
quark to hadron transverse momentum a Lorentz transformation leaving $k^-$
and $P_1^-$ unchanged needs to be performed. The fragmentation
functions are real and in fact, depend on $z$ and $\bm{k}_T^\prime{}^2$ only.

We note that after integration over $\bm k_T^{}$ several functions
disappear. In the case of $\text{Tr} (\Delta \, i\sigma^{i-}\gamma_5)$ a 
specific combination remains, 
namely $H_1$ $\equiv$ $H_{1T} + (\bm k_T^2/2M_1^2)\,H_{1T}^\perp$.

The
choice of factors in the definition of fragmentation functions is such that
$\int dz\,d^2\bm k_T^\prime\,D_1(z,\bm k_T^\prime) = N_h$, where $N_h$ is 
the number of produced hadrons.

Note that the decay probability for an unpolarized quark with non-zero 
transverse momentum can lead to a transverse polarization in the production 
of spin-1/2 particles. This polarization is orthogonal to the quark transverse
momentum and the probability is given by the function $D_{1T}^{\perp}$. 
In the same way, oppositely transversely polarized quarks with
non-zero transverse momentum can produce unpolarized hadrons or 
spinless particles, with different probabilities. In other words: there can be
a preference for one or the other transverse polarization direction of the 
quark
(aligned or opposite relative to its transverse momentum) to fragment into an
unpolarized hadron. This difference is described
by the function $H_1^\perp$. It is the one appearing in the so-called
Collins effect \cite{Collins-93b}, which predicts a single transverse 
spin asymmetry in for instance semi-inclusive DIS, and arises due to intrinsic
transverse momentum.  

The functions $D_{1T}^{\perp}$ and $H_1^\perp$ are what are 
generally called `time-reversal odd' functions. For a discussion on the
meaning of this, we refer to Ref.\ \cite{Boer} and earlier references 
\cite{DeRujula-71}
; here we only remark that it
does not signal a violation of time-reversal invariance of the theory, but
rather the presence of final state interactions.

The cross sections are obtained from the hadron tensor after contraction with
the lepton tensor
\begin{eqnarray}
L^{\mu \nu} & = & (g_V^l{}^2 + g_A^l{}^2) \,Q^2\, \Biggl[
- \left( 1 - 2y + 2y^2 \right) g_\perp^{\mu \nu}
+ 4y(1-y) \hat z^\mu \hat z^\nu
\nonumber \\ && \qquad \qquad
-4y(1-y)\left( \hat l_\perp^\mu\hat l_\perp^\nu +\frac{1}{2}\,g_\perp^{\mu \nu}
\right)
- 2(1 - 2y)\sqrt{y(1-y)}\,\,\hat z_\ph^{\{ \mu}\hat l_\perp^{\nu \}} \Biggr]
\nonumber \\ &&\quad - (2 g_V^l g_A^l) \,Q^2\, \Biggl[
+i\,(1-2y)\,\epsilon_\perp^{\mu \nu}
- 2i \,\sqrt{y(1-y)}\,\,\hat l{}_{\perp\rho} 
\epsilon_\perp^{\rho\,[ \mu}
\hat z_\ph^{\nu ]} \Biggr],
\end{eqnarray}
where $\{ \mu\nu \}$ indicates symmetrization of indices, $[ \mu\nu ]$ 
indicates antisymmetrization. The fraction $y$ is defined to be $y=
P_2 \cdot l/ P_2 \cdot q \approx l^-/q^-$, which in the lepton center of mass 
frame equals 
$y=(1 + \cos \theta_2)/2$, where 
$\theta_2$ is the angle of hadron two with respect to the momentum of the 
incoming leptons.

Azimuthal angles inside the perpendicular plane are defined with respect 
to $\hat l_\perp^\mu$, defined to be the normalized perpendicular part of the 
lepton momentum $l$, $\hat l_\perp^\mu = l_\perp^\mu /(Q \sqrt{y(1-y)})$:
\begin{eqnarray}
\hat{l}_\perp \cdot a_\perp &=& - |{\bm a}_\perp | \cos \phi_a,
\label{anglecos} \\
\epsilon^{\mu\nu}_\perp \hat{l}_{\perp\mu} a_{\perp\nu}&=& |{\bm a}_\perp | 
\sin \phi_a,
\label{anglesin}
\end{eqnarray}
for a generic vector $a$.

The vector and axial couplings to the $Z$ boson are given by:
\begin{eqnarray} 
g_V^j &=& T_3^j - 2 \, Q^j\,\sin^2 \theta_W,\\
g_A^j &=& T_3^j,
\end{eqnarray} 
where $Q^j$ denotes the charge and $T_3^j$ the weak isospin of 
particle $j$ (i.e., $T_3^j=+1/2$ for $j=u$ and $T_3^j=-1/2$ for
$j=e^-,d,s$). Combinations of the couplings occurring frequently in the
formulas are 
\begin{eqnarray} 
c_1^j &=&\left(g_V^j{}^2 + g_A^j{}^2 \right),
\nonumber\\[2 mm]
c_2^j &=&\left(g_V^j{}^2 - g_A^j{}^2 \right),
\qquad\qquad j=\ell\;\;\mbox{or}\;\;u,d,s
\nonumber\\[2 mm]
c_3^j &=&2 g_V^j g_A^j.
\end{eqnarray}
As well, we will use the following kinematical factors:
\begin{eqnarray}
A(y) &=& \left(\frac{1}{2}-y+y^2\right) \ \stackrel{cm}{=} 
\ \left( 1 + \cos^2\theta_2 \right)/4,
\nonumber\\
B(y) &=& y\,(1-y) \ \stackrel{cm}{=} \ \sin^2 \theta_2/4,
\nonumber\\[2 mm]
C(y) &=& (1-2y) \ \stackrel{cm}{=} \ - \cos \theta_2.
\end{eqnarray}
We obtain in leading order in $1/Q$ and $\alpha_s$ the following 
expression for the cross section in case of unpolarized 
(or spinless) final state hadrons:
\begin{eqnarray}
\lefteqn{\frac{d\sigma (e^+e^-\to h_1h_2X)}{d\Omega dz_1 dz_2 
d^2{\bm q_T^{}}}=
\sum_{a,\overline a}\;\frac{3\,\alpha_w^2\, Q^2}
{(Q^2 - M_Z^2)^2+\Gamma_Z^2M_Z^2}
\;z_1^2z_2^2\;\Bigg\{ 
          \left(
          c_1^\ell\,c_1^a\;A(y)-{\scriptstyle \frac{1}{2}}\, 
c_3^\ell\,c_3^a C(y)\right) 
                {\cal F}\left[D_1^a\overline D{}_1^a\right]}
\nonumber\\ && 
       \hspace{48mm} 
        + \cos(2\phi_1)\;c_1^\ell\,c_2^a\;B(y)\;
             {\cal F}\left[\left(2\,\hat{\bm{h}}\!\cdot\!
                           \bm k_T^{}\,\,\hat{\bm{h}}\!\cdot\!\bm p_T^{}\,
                        -\,\bm k_T^{}\!\cdot \!\bm p_T^{}\,\right)
                    \frac{H_1^{\perp a}\overline H{}_1^{\perp a}}{M_1M_2}
\right]
\Bigg\},
\label{LO-OOOnopol}
\end{eqnarray}
where 
$d\Omega$ = $2dy\,d\phi^l$ and $\phi^l$ gives the 
orientation of $\hat l_\perp^\mu$. We use the convolution notation
\begin{equation} 
{\cal F}\left[D^a\overline D{}^a\, \right]\equiv
\int d^2\bm k_T^{}\; d^2\bm p_T^{}\;
\delta^2 (\bm p_T^{}+\bm k_T^{}-\bm 
q_T^{})  D^a(z_{1},z_{1}^2 \bm{k}_T^2) 
\overline D{}^a(z_{2},z_{2}^2 \bm{p}_T^2).
\end{equation}
The angle $\phi_1$ is the azimuthal angle of $\hat h$ 
(see Fig.\ \ref{fig:kinann}). In order to deconvolute these expressions we can
define weighted cross sections
\begin{equation}
\left< W \right>_{A}=
\int \frac{d\phi^\ell}{2\pi}\,d^2\bm q_T^{} 
\ W\,\frac{d\sigma (e^+e^-\to h_1h_2X)}{d\Omega dz_1 dz_2 
d^2{\bm q_T^{}}},
\end{equation}
where $W$ = $W(Q_T,\phi_1,\phi_2,\phi_{S_1},\phi_{S_2})$. The subscript
$A$ denotes the polarization in the final state for hadron one, 
with as possibilities unpolarized ($O$, including the case of 
summation over spin), longitudinally polarized
($L$) or transversely polarized ($T$). We postpone the discussion of 
the additional structures and information accessible by 
measuring the polarization of one of the final state hadrons 
to the end of this letter. 

Even without determining polarization of a final state hadron a subtle test
of our understanding of spin transfer mechanisms in perturbative QCD 
can be done. The information on the production of a transversely polarized 
quark-antiquark pair, which subsequently fragment into unpolarized (or
spinless) hadrons with probabilities depending on the
orientation of the (anti)quark's spin vector relative to its 
transverse momentum, 
is contained in the $\cos(2\phi_1)$ azimuthal asymmetry\footnote{This 
asymmetry is
not to be confused with the $\cos(2\phi)$
asymmetry found by Berger \cite{Berger-80}, which is $1/Q^2$ suppressed.}.
To access this information we utilize the weighted cross sections
\begin{equation}
\left< 1 \right>_{O} 
\ = \ 
\frac{3\,\alpha_w^2\, Q^2}
{(Q^2 - M_Z^2)^2+\Gamma_Z^2M_Z^2}
\sum_{a,\bar a}\; 
         \left(c_1^\ell\,c_1^a\;A(y)-{\scriptstyle \frac{1}{2}}\,
c_3^\ell\,c_3^a C(y)\right) 
         \;D_1^a(z_1)\,\overline D{}_1^a(z_2),
\label{wXsection-1_O}
\end{equation}
\begin{equation}
\left< \frac{Q_T^2}{4 M_1M_2}\,\cos(2\phi_1) \right>_{O} 
\ = \ 
\frac{3\,\alpha_w^2\,Q^2}
{(Q^2 - M_Z^2)^2+\Gamma_Z^2M_Z^2}
\sum_{a,\bar a}\; 
c_1^\ell\,c_2^a\;B(y)\;H_1^{\perp(1)a}(z_1)
\,\overline H_1^{\perp (1)a}(z_2),
\label{wXsection-cos2phi_O}
\end{equation}
where the $\bm{k}_{T}^2$-moments for a generic fragmentation function $F$ are 
defined by
\begin{equation}
F^{(n)}(z_{i})= z_{i}^2 \int d^2 \bm{k}_{\scriptscriptstyle T}\,
\left(\frac{\bm{k}_{\scriptscriptstyle T}^2}{2 M_i^2}\right)^n\;  
F(z_{i},z_{i}^2 \bm{k}_{\scriptscriptstyle T}^2).
\end{equation}

We now like to focus on the weighted cross section
defined in Eq.~(\ref{wXsection-cos2phi_O}) and discuss its possible 
measurement. 
In order to be able to observe the $\cos(2\phi)$ dependence one must look at
two jet events in unpolarized electron-positron scattering. In each jet one
identifies a fast hadron with momentum fractions $z_1$ and $z_2$ 
respectively. One of the hadrons (say two) together with the leptons 
determines the lepton scattering plane as is indicated in Fig.\ 
\ref{fig:kinann}.
In the lepton CM system hadron two determines the $\hat{z}$-direction with 
respect to which the azimuthal angles are measured. One needs in 
particular the azimuthal angle $\phi_1$ of the other hadron (one) as well 
as its transverse momentum $\bm P_{1\perp}$, which determines 
$Q_T$ = $\vert \bm P_{1\perp}\vert/z_1$. The $\cos(2\phi)$ angular 
dependence then can be analyzed by calculating the weighted cross section of
Eq.~(\ref{wXsection-cos2phi_O}). 

\begin{figure}[htb]
\begin{center}
\leavevmode \epsfxsize=8cm \epsfbox{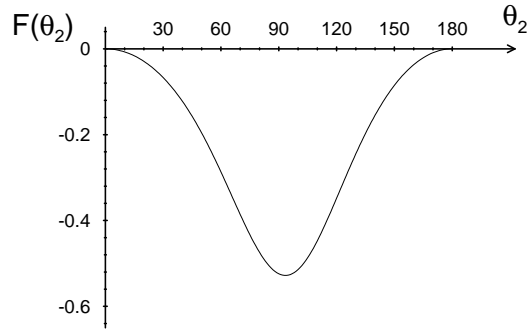}
\caption{\label{fig:factor} Factor defined in Eq.~(\ref{factor}) depending on
the center of mass angle $\theta_2$.}
\end{center}
\end{figure}

For an order of magnitude estimate, we
consider the situation of the produced hadrons being a $\pi^+$ and a
$\pi^-$. Furthermore, we assume $D_1^{u\to\pi^+}(z)=D_1^{\bar d\to\pi^+}(z)$
($D_1^{d\to\pi^-}(z)=D_1^{\bar u\to\pi^-}(z)$, respectively) and 
neglect unfavored
fragmentation functions like $D^{d\to\pi^+}(z)$ etc.; and similar
for the time-reversal odd functions. The equalities for the $D_1$ functions
seem quite safe on grounds of isospin and charge conjugation, the same
assumptions might be non-trivial for the $H_1^\perp$ functions. As a
consequence of these assumptions the fragmentation
functions can be taken outside the flavor summation, and we obtain
\begin{equation}
\left< \frac{Q_T^2}{4M_1M_2}\,\cos(2\phi_1) \right>_{O}
= F(y)
\,\frac{H_1^{\perp(1)}(z_1)}{D_1(z_1)}
\,\frac{H_1^{\perp(1)}(z_2)}{D_1(z_2)}
\,\left< 1 \right>_{O},
\label{ratio}
\end{equation}
where
\begin{equation}
F(y)=
\frac{\sum_{a=u,\bar d}\;c_1^e\,c_2^a\;B(y)}{\sum_{a=u,\bar d}\;
\left(c_1^e\,c_1^a\;A(y)-{\scriptstyle \frac{1}{2}}
\,c_3^e\,c_3^a\,C(y)\right)}.
\label{factor}
\end{equation}
This factor is shown in in Fig.~\ref{fig:factor} as a function of the
center of mass angle $\theta_2$ (we use $\sin^2\theta_W=0.2315$~\cite{PDG}). 
At an angle close to $90^o$ we observe the largest effect.
In order to get an estimate of the true asymmetry at the level of
count rates, one should compare Eq.~(\ref{wXsection-cos2phi_O}) 
with the weighted cross section $\left< Q_T^2/4M_1M_2 \right>_{O}$. 
To estimate the ratio of those two quantities, we use
as argued in reference \cite{Kotzinian-Mulders-97}, for the 
ratio of the 
fragmentation functions $H_1^{\perp(1)}(z_1)/D_1(z_1) ={\cal O} (1)$,
although this is likely an optimistic estimate. 
From the average transverse momentum squared of produced 
pions in one jet, for which we take 0.5 (GeV/c)$^2$ \cite{Exptrans}, one  
obtains an estimate for the average transverse momentum of pions in jet one 
with respect to a given pion in jet two. This leads at $z_1=z_2=1/2$ to 
$\left< Q_T^2/4M_\pi^2 \right>_{O}\approx 50 \,\left< 1 \right>_{O}$
and consequently to an estimate at the percent level for the ratio 
$\left<\left(Q_T^2/4M_\pi^2\right) \,\cos(2\phi_1) \right>_{O}
/\left<Q_T^2/4M_\pi^2 \right>_{O}$.
Such an azimuthal dependence in the unpolarized cross section, 
however, may be detectable in present-day electron-positron 
scattering experiments.

The situation where hadron two 
is taken to be a jet, which in this back-to-back jet situation is equivalent 
to analyzing the azimuthal structure of hadrons inside a jet, is obtained by 
considering
$\overline D{}_1^a(z_2) = \delta(1-z_2)$ and 
$\overline H{}_1^{\perp (1)a}(z_2) = 0$. This gives the familiar result
for $\int dz_2\,\left< 1\right>_O$ and it gives zero for the $z_2$-integrated
$\cos 2\phi$ azimuthal asymmetry.

The experimental determination of the polarization of (one of) the final state
hadrons offers further opportunities to reveal the hadronic structure in terms
of spin-dependent fragmentation functions. We assume in the following that 
the spin vector of hadron one, i.e.~$S_1$, is known (reconstructed), having
in mind the example of a produced $\Lambda$ and its self-analyzing
properties. 
We observe a rich structure of angular dependences due to polarization.

Again, weighted cross sections are the appropriate means to separate out
specific functions. For instance, the weighted cross section
\begin{equation}
\left< \frac{Q_T}{M_2}\,\sin(\phi_1+\phi_{S_1}) \right>_{T} 
\ = \ 
\vert \bm S_{1T}\vert\,\frac{3\,\alpha_w^2\,Q^2}
{(Q^2 - M_Z^2)^2+\Gamma_Z^2M_Z^2}
\sum_{a,\bar a}\;c_1^\ell\,c_2^a\;B(y)\;
\,H_{1}^{a}(z_1)\,\overline H{}_1^{\perp (1)a}(z_2)
\end{equation}
picks out 
the term which is the closest analogue to the Collins effect
\cite{Collins-93b} in semi-inclusive lepton-hadron scattering 
\cite{Mulders-Tangerman-96}.
We note that a confirmation of the $\cos(2\phi)$ asymmetry, 
also implies a confirmation of the Collins effect. A complete
list of weighted cross sections at leading order is given
in Table \ref{tabel}.

\begin{table}[htb]
\caption{\label{tabel}Weighted cross sections for $S_1 \neq 0, S_2 = 0$}
\begin{tabular}{cll}
& & \\
$W$ & $A$
& $\left<W\right>_A \cdot \left[3\,\alpha_w^2\,Q^2 \big/ \left( 
(Q^2 - M_Z^2)^2+\Gamma_Z^2M_Z^2 \right) \right]^{-1} $\\
& \\ \hline
& & \\
$1$ & $L$ & $-\lambda_1\,\sum_{a,\bar a}\;
        \left(c_1^\ell\,c_3^a\;A(y)- {\scriptstyle \frac{1}{2}} \, 
c_3^\ell\,c_1^a\;C(y) \right)\;
        G_{1L}^a(z_1)\,\overline D{}_1^a(z_2)$ \\
& & \\
$(Q_T^2/4 M_1M_2)\,\sin(2\phi_1)$ & $L$ & $\lambda_1\,
\sum_{a,\bar a}\;c_1^\ell\,c_2^a\;B(y)\;
         H_{1L}^{\perp(1)a}(z_1)\,\overline H{}_1^{\perp (1)a}(z_2)$ \\
& & \\
$(Q_T/ M_1)\,\sin(\phi_1-\phi_{S_1})$ & $T$ & $\vert \bm S_{1T}\vert\,
\sum_{a,\bar a}\;
        \left(c_1^\ell\,c_1^a\;A(y)-{\scriptstyle \frac{1}{2}} \, 
c_3^\ell\,c_3^a\;C(y)\right)
        \;D_{1T}^{\perp(1)a}(z_1)\,\overline D{}_1^a(z_2)$ \\
& & \\
$(Q_T/M_2)\,\sin(\phi_1+\phi_{S_1})$ & $T$ & $-\vert \bm S_{1T}\vert\,
\sum_{a,\bar a}\;c_1^\ell\,c_2^a\;B(y)\;
\,H_{1}^{a}(z_1)\,\overline H{}_1^{\perp (1)a}(z_2)$ \\
& & \\
$(Q_T^3/6M_1^2M_2)\,\sin(3\phi_1 - \phi_{S_1})$ & $T$ & $-\vert \bm
S_{1T}\vert\, 
\sum_{a,\bar a}\;c_1^\ell\,c_2^a\;B(y)\;
H_{1T}^{\perp(2)a}(z_1)\,\overline H{}_1^{\perp (1)a}(z_2)$ \\
& & \\
$(Q_T/M_1)\,\cos(\phi_1-\phi_{S_1})$ & $T$ & $\vert \bm S_{1T}\vert\,
\sum_{a,\bar a}\;
        \left(c_1^\ell\,c_3^a\;A(y)\;-{\scriptstyle \frac{1}{2}} \,
c_3^\ell\,c_1^a\;C(y)\right)
        \;G_{1T}^{(1)a}(z_1)\,\overline D{}_1^a(z_2)$ \\
& & \\
\end{tabular}
\end{table}

In conclusion, we have presented the leading asymmetries in 
inclusive two-hadron production in electron-positron annihilation at the 
$Z$ pole. We have investigated unpolarized and single spin asymmetries. 
We included the effects of intrinsic transverse momentum and in this sense
our results are an extension of those of Ref.\ \cite{Chen-et-al-95}.   
The azimuthal dependence in the unpolarized differential cross section 
is a $\cos(2\phi)$ asymmetry, which arises 
solely due to the intrinsic transverse momenta of the quarks. 
An extensive discussion on how to measure 
this asymmetry and the accompanying time-reversal odd fragmentation 
functions is given. A simple estimate 
indicates that the asymmetry could be at the percent level, hence
it can perhaps be observed in present-day electron-positron scattering
experiments. In confirming the existence of this asymmetry one also 
confirms the Collins effect, without the need of a polarization
measurement.

{\em Note added:} In a preliminary study \cite{Delphi} a similar correlation
in back-to-back jets was already experimentally investigated.
We find that it involves moments of the functions $H_1^\perp$ and
$\overline H{}_1^\perp$, different from the ones in our correlation.
In this study no significant result was found using the 1991 to 1994
LEP data. In this analysis {\em three} momenta in the final state need
to be determined, namely besides two hadron momenta also the jet axis,
and hence there are two azimuthal angles, $\phi$ and $\phi^\prime$,
yielding a $\cos(\phi+\phi^\prime)$ asymmetry. 

\acknowledgments 
We like to thank Dani\"{e}l van Dierendonck and Niels Kjaer for useful
discussions on the experimental aspects. Furthermore, we 
thank Thierry Gousset and Oleg Teryaev for several discussions. 
This work is part of the research program of the foundation for 
Fundamental Research of Matter (FOM), the National Organization 
for Scientific Research (NWO) and the TMR program ERB FMRX-CT96-0008.


\begin{thebibliography}{30}

\bibitem{Boer} 
D. Boer, R. Jakob and P.J. Mulders, Nucl.~Phys.~B 504 (1997) 345.

\bibitem{Chen-et-al-95}
K. Chen, G.R. Goldstein, R.L. Jaffe and X. Ji, Nucl.~Phys.~B 445 (1995) 380. 

\bibitem{Artru-Collins-96}
X. Artru and J.C. Collins, Z.~Phys.~C 69 (1996) 277.
 
\bibitem{Collins-Soper-82}
J.C. Collins and D.E. Soper, Nucl.~Phys.~B 194 (1982) 445.

\bibitem{Collins-93b}
J.C. Collins, Nucl.~Phys.~B 396 (1993) 161.

\bibitem{DeRujula-71}
A. de Rujula, J.M. Kaplan, and E. de Rafael, Nucl.~Phys.~B 35 (1971) 365;\\
K.~Hagiwara, K.~Hikasa, N.~Kai, Phys.~Rev.~D 27 (1983) 84;\\
D. Atwood, G. Eilam and A. Soni, Phys.~Rev.~Lett.~71 (1993) 492.

\bibitem{Berger-80}
E.L. Berger, Z.~Phys.~C 4 (1980) 289; Phys.~Lett.~B 89 (1980) 241. 

\bibitem{PDG} Particle Data Group, Phys.~Rev.~D 54 (1996) 1. 

\bibitem{Kotzinian-Mulders-97}
A.M. Kotzinian and P.J. Mulders, Phys.~Lett.~B 406 (1997) 373.

\bibitem{Exptrans}
EMC Collaboration, M.~Arneodo {\it et al.}, Z.~Phys.~C 34 (1987) 277;\\
EMC Collaboration, J. Ashman {\it et al.}, Z.~Phys.~C 52 (1991) 361;\\
E665 Collaboration, M.R. Adams {\it et al.}, Phys.~Rev.~D 48 (1993) 5057.

\bibitem{Mulders-Tangerman-96}
P.J. Mulders and R.D. Tangerman, Nucl.~Phys.~B 461 (1996) 197;
Nucl.~Phys.~B 484 (1997) 538 (E).

\bibitem{Delphi}
DELPHI Collaboration, W. Bonivento {\it et al.}, Internal Note DELPHI-95-81
PHYS 516, Contribution eps0549 to the EPS-HEP 95 conference, Brussels
(1995), unpublished.

\end{thebibliography}
\end{document}